\documentclass[journal]{IEEEtran}
\usepackage[colorlinks=true,citecolor=blue,pdfborder={1 1 1}, pdfborderstyle={/S/U/W 1}, pdftex]{hyperref}
\usepackage{cite}
\usepackage[pdftex]{graphicx}
\usepackage[cmex10]{amsmath}
\usepackage{mdwmath}
\usepackage{array,booktabs,siunitx,xcolor,colortbl}
\newcolumntype{C}[1]{>{\centering\arraybackslash}m{#1}}%
\usepackage{flushend}

\begin{document}

\title{Impact of Filling Ratio on Subwavelength Optical Imaging with Two Different Geometries}

\author{Md.~Ibrahim~Khalil,
        Md.~Saad-Bin-Alam,
        Atiqur~Rahman,~\IEEEmembership{Member,~IEEE,}
        and~Pavel~A.~Belov,~\IEEEmembership{Member,~IEEE}
\thanks{Md.~Ibrahim~Khalil, Md.~Saad-Bin-Alam and Atiqur~Rahman are with the Department of Electrical Engineering and Computer Science, North South University, Dhaka-1229, Bangladesh. (e-mail: \href{ibrahim030@eecs.northsouth.edu}{ibrahim030@eecs.northsouth.edu}, \href{saadbinalam.nsu.bd@gmail.com}{saadbinalam.nsu.bd@gmail.com} and \href{atiqur@northsouth.edu}{atiqur@northsouth.edu}).}
\thanks{Pavel~A.~Belov is with the Department of Photonics and Optoinformatics, St. Petersburg National Research University of Information Technologies, Mechanics and Optics (IMTO), Kronverkskiy pr., 49, 197101 St. Petersburg, Russia. (e-mail:\href{belov@phoi.ifmo.ru}{belov@phoi.ifmo.ru})}
\thanks{Atiqur~Rahman and Pavel~A.~Belov  are also with School of Electronic Engineering and Computer Science, Queen Mary University of London, Mile End Road, London E1 4NS, United Kingdom}
\thanks{Manuscript received April 09, 2014; revised --- }}



\maketitle

\begin{abstract}
Metallic nano-structured lens has strong potential application in transporting subwavelength image information. This be achieved by manipulating the length of the nanorod and the periodicity of the rod array. In this paper, we have demonstrated the impact of filling ratio on the subwavelength imaging capabilities of such lenses. Through full-wave electromagnetic simulation, we have demonstrated that the imaging performance of silver (Ag) nanorod array does not only depend on the length and periodicity but also on the filling ratios or the radius of the wire medium. We have investigated this impact in two different geometries for nanorod array in the form of cylindrical and triangular rod and also examined their performances for different filling ratios.
\end{abstract}


\section{Introduction}
\PARstart{I}{maging} devices based on conventional lenses cannot offer resolution better than half of the operating wavelength due to diffraction limit as well as Rayleigh criterion of half of the operating wavelength \cite{Ref1}. Imaging below the half wavelength resolution ($\leq \lambda/2$) or in subwavelength scale requires capturing evanescent waves which vanishes exponentially as it travels away from the source and requires time-consuming point-by-point scanning by near-field scanning microscopy \cite{Pavel_Nature,Pavel_2,Ref3}. Thus, in order to overcome these restrictions, several imaging techniques were proposed in last decades including perfect lens, silver super-lens, hyper-lens, stimulated emission depletion (STED) fluorescence microscope etc. \cite{Pavel_Nature,Pavel_2,Ref3,Ref4,Ref5,Ref6,Ref7}. Besides these, metallic wire media as subwavelength imaging system was also proposed in \cite{Ref8,Ref9}. However, a complete procedure to model such a metallic nanowire based lens for imaging in the visible range was discussed in \cite{Ref10,Ref11,Ref12}.\\

\begin{figure}[htb]
\centerline{\includegraphics[width=8.70cm]{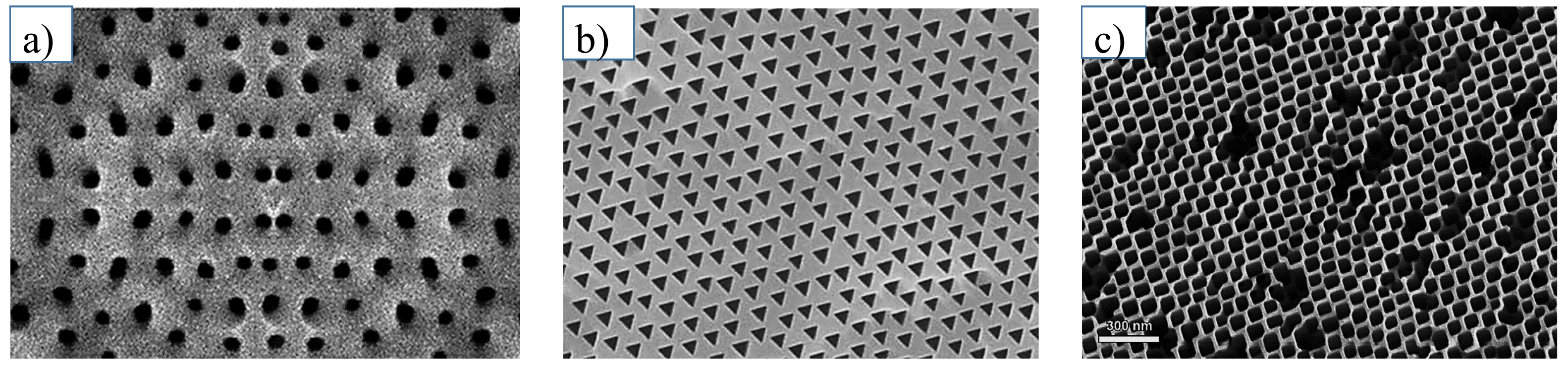}}
\caption{Porous matrices for fabrication of optical nanowire arrays (a) circular, (b) triangular and  (c) square cross-sections [\textcolor{blue}{15}]-[\textcolor{blue}{17}]}
\label{Fig:1}
\end{figure}

In the case of subwavelength imaging at optical frequency, the main focus was on manipulating the length of nanowire array ($l$) to improve resolution of the image \cite{Ref10,Ref11}. Even though, the concept of finding optimal parameters to achieve sub-wavelength imaging was investigated by optimizing the length of the rod array \cite{Ref12}. Kosulnikov \emph{et. al.} drew attention to the effect of filling factors (defined as $f_r \approx \pi(r/a)^2$, where,`$r$' is the radius of the nanorods and `$a$' is the lattice constant) at infrared range \cite{Ref13}. These types of investigations are even more necessary in the optical range since it is very difficult to maintain accurate dimensions of nanowire array as the radius of the wire medium is much smaller than the skin depth ($\delta$). Thus in the context of practical realization, where the unit is counted in nanometer, the values of optimal length and radius ensure the ultimate limit of error in the manufacturing process. In this paper, we have calculated the optimal filling ratios (ratio of the rod cross-sectional area with the lattice cross-sectional area) and hence the range of radius allowed to achieve imaging. Through full wave electromagnetic simulations we have established our findings using infinitesimal silver nanorod array in equilateral triangular or, hexagonal lattice arrangement with two different cross-sectional geometries. Nanorod array with cylindrical \cite{Ref17}, triangular \cite{Ref18} and square \cite{Ref19} porous matrices for fabrication was realized using different substrate material as shown in Fig. \ref{Fig:1} which can be employed for experimental demonstrations in the near future.

\begin{figure}[htb]
\centerline{\includegraphics[width=8.70cm]{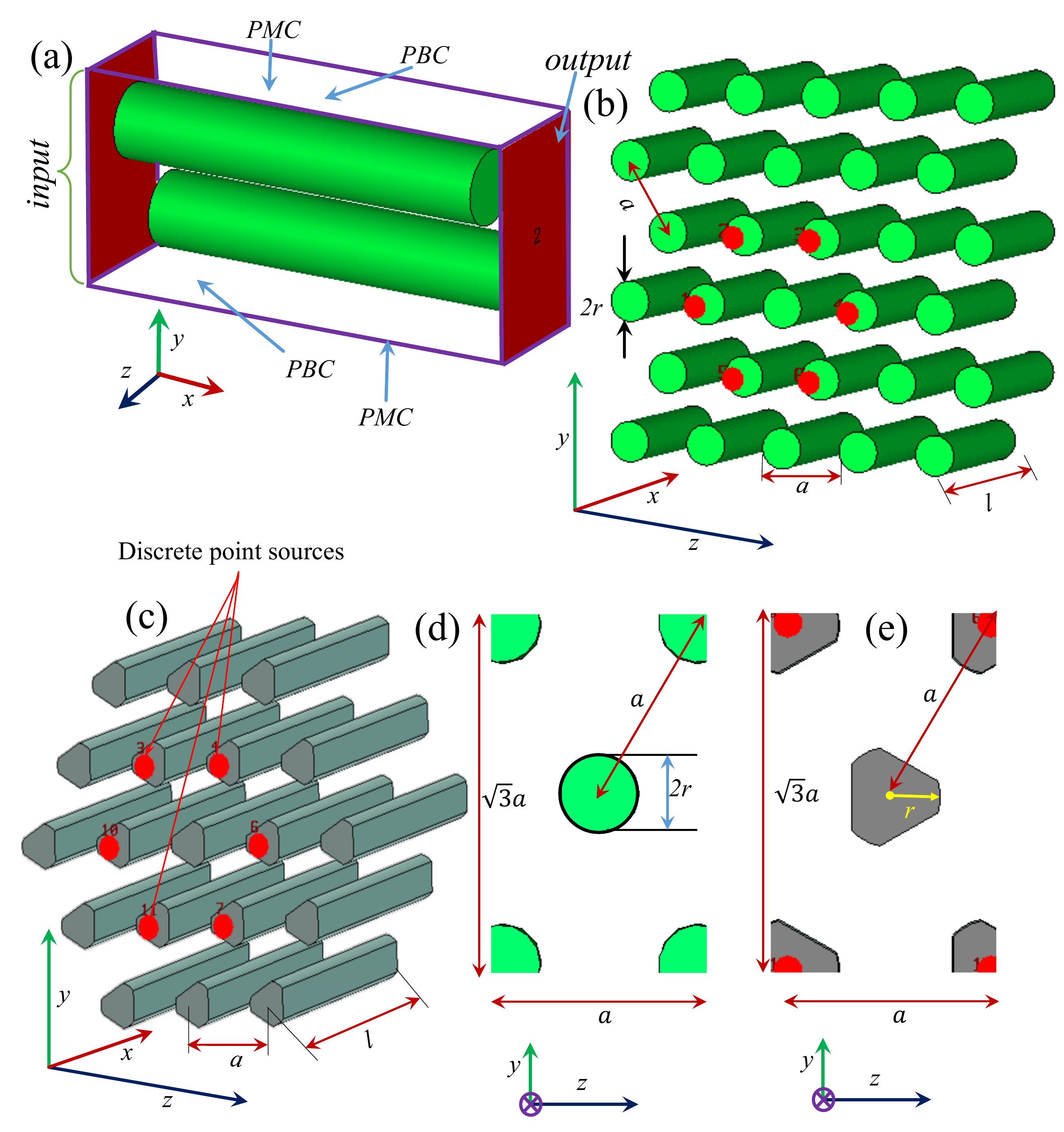}}
\caption{Schematic diagram of wire medium: (a) unit cell employed to calculate the transmission and reflection coefficients of an infinite array having triangular lattice, (b, c) three dimensional nanorod array modeled in CST\textsuperscript{TM} Studio Suite where cylindrical as well as triangular shaped nanorod was employed (d, e) the cross-sectional diagram for both the nanorod. Notation: PMC, perfect magnetic conductor boundaries; PBC, periodic boundary conditions.}
\label{Fig:2}
\end{figure}

 In our analysis, we have considered filling ratio, $f_r\approx(2\pi/3\sqrt{3})(r/a)^2$ for cylindrical shaped wire array and $f_r\approx(1/2)(r/a)^2$ for triangular shaped wire array. Here the parameter `$a$' is known as periodicity or lattice constant of the array and `$r$' is the radius of circular cross-section of cylindrical shape Fig.\ref{Fig:2}(d) and also represents the distance of the polished edged equilateral triangular cross-section Fig.\ref{Fig:2}(e). In this manner we compute the allowable filling ratios or radius for both cross-sectional geometries and found triangular rod array offers higher filling ratio tolerances than the cylindrical one while other parameters like periodicity `$a$' and length `$l$' were kept unchanged. 
 
\section{Analytical modeling of nanorod array}
The schematic view of the unit cell and silver nanorod array with triangular lattice arrangement are shown in Fig. \ref{Fig:2} which has been numerically modeled using commercial full-wave electromagnetic simulator CST\textsuperscript{TM} Studio Suite \cite{Ref14}. The rod length is set to be 80nm for both the geometries with the same periodicity of 40nm. We have used Drude Dispersion Model for the permittivity of \emph{Ag} which was defined as $\epsilon_m (\omega) = \epsilon_\infty -\omega_p^2/(\omega^2+j \gamma \omega)$ where, $\epsilon_\infty$ = 4.9638, $\omega_p = 1.4497 \times 10^{16}$ rad/s and $\gamma = 8.33689 \times 10^{13}$/s \cite{Ref15, Ref16}. \\

\section{Imaging with cylindrical nanorod array}
We have investigated the imaging capabilities of this device with discrete coherent source which were formed by six in-phase point sources polarized along the rod length ($x$-axis) and placed on the rod axis forming a hexagon, as shown in Fig. \ref{Fig:2} (b, c). The importance of using such a source was discussed in \cite{Ref11,Ref12}. 

\begin{figure}[htb]
\centerline{\includegraphics[width=8.70cm]{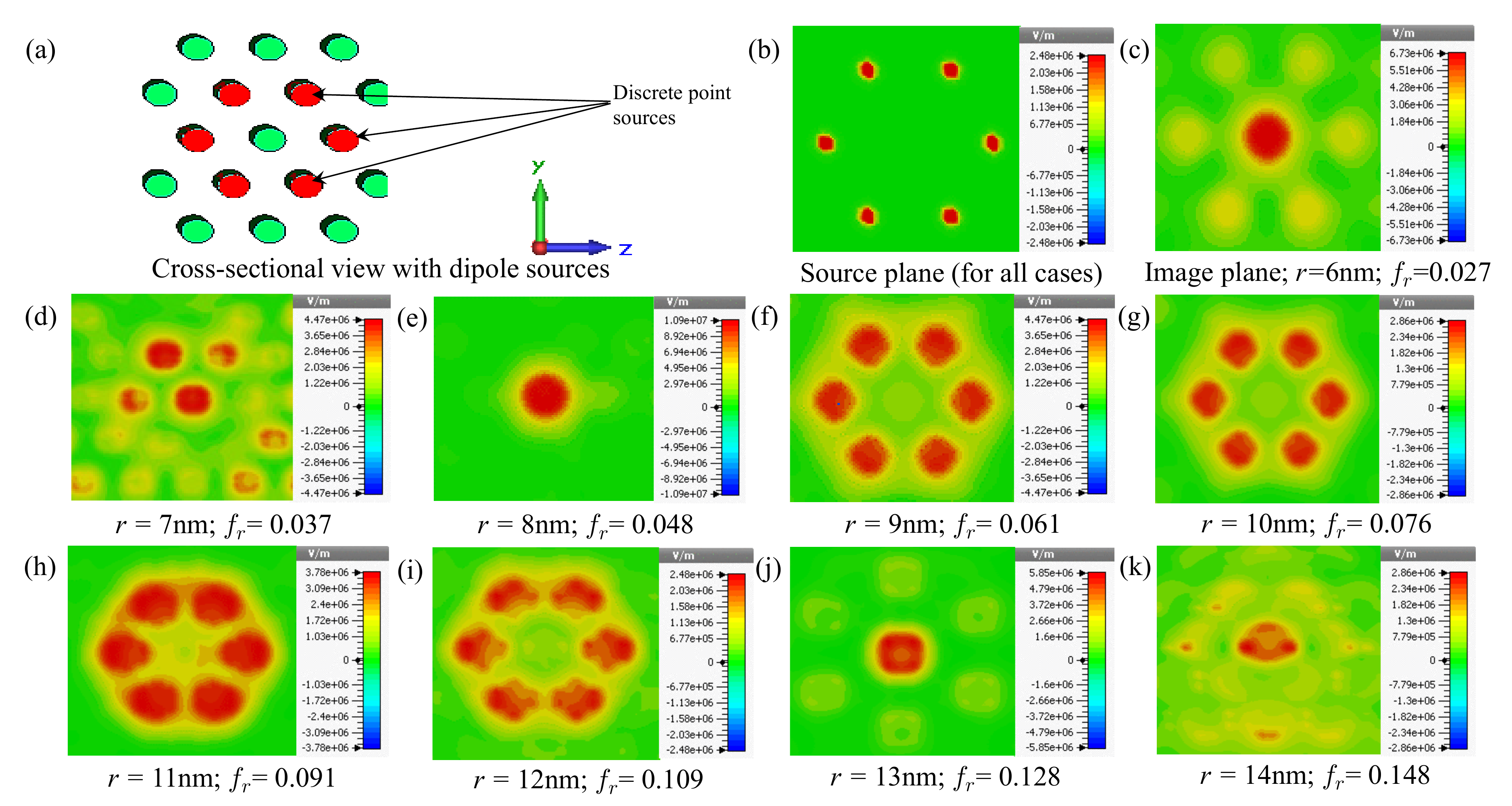}}
\caption{Source plane and image plane field distributions at 614THz for various radii of cylindrical shaped rod array ($0.15a$ to $0.35a$) of the rod, (a) hexagonal point source (b) Source Plane and (c-k) image plane for $r$=6-14nm respectively.}
\label{Fig:3}
\end{figure}

In order to show the impact of filling ratio, we have examined the $E$-field distributions calculated at the image plane (10nm away from the back interface) at 614THz for a range of filling ratios ($0.027\leq f_r \leq 0.148$ or, $6nm \leq r \leq 14nm$) as shown in Fig. \ref{Fig:3}. As observed from Fig. \ref{Fig:3} the radii, \emph{r} = 9, 10, 11 and 12nm, the image plane field distributions matches with the one obtained in the source plane. However for other values of `$r$' field distributions at the image plane gives us misleading views of the source plane. It is evident from the figure that for a rod radius larger than 12nm the image plane field distribution gives an impression of single point source located at the center of the hexagon, whereas for a rod radius smaller than 9nm extremely inconsistent images are produced.\\ 

\section{Transmission \& reflection characteristics of cylindrical nanorod array}
In order to validate the results obtained in the previous section we have calculated the transmission coefficients of such an array applying transverse magnetic (TM) polarization. The architecture of the unit cell is illustrated in Fig. \ref{Fig:2}(a) for radii (filling ratio) between $6nm \leq r \leq 15nm$. Here periodic boundary conditions have been applied in both $y$ and $z$ directions. The length `$l$' of \emph{Ag} nanowire has been fixed at 80nm as per Ref. \cite{Ref11,Ref12} because they successfully proved that the best image is achieved at 614THz corresponds to the aforementioned rod length. The periodicity `$a$' of the nanorod array has been set to a constant value of 40nm as in \cite{Ref11,Ref12}. Transmission and reflection coefficients for various filling ratio, `$f_r$' have been calculated by changing the radius from 6nm to 15nm with an increment of 1nm.\\

\begin{figure}[htb]
\centerline{\includegraphics[width=8.70cm]{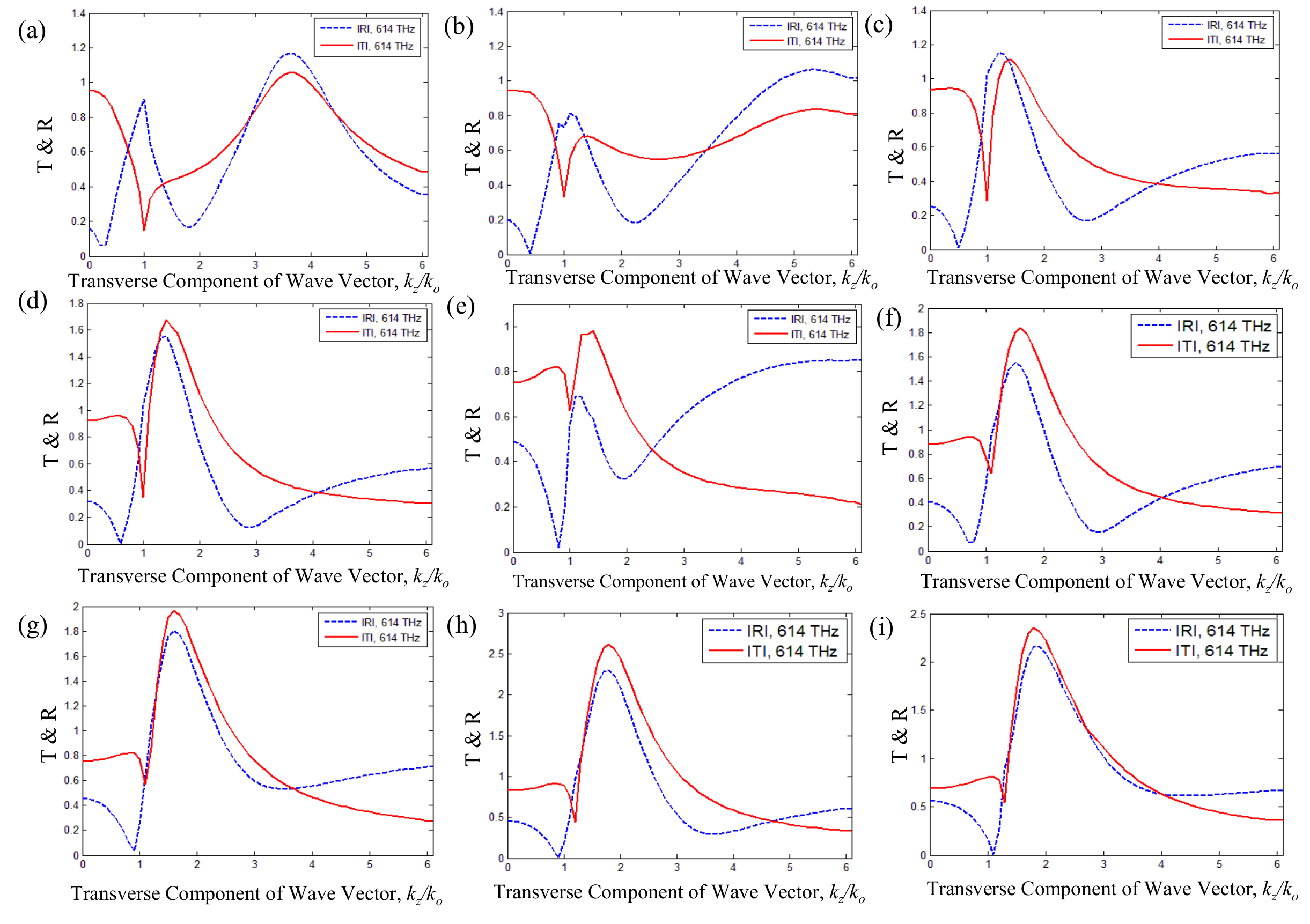}}
\caption{Transmission Coefficient (TM Modes) patterns for different radius of Ag rod at an operating wavelength of 614THz. In each case the length of each rod and the periodicity are kept same as the Ref. [\textcolor{blue}{12, 13}] while the radius varied from 6nm to 14nm with 1nm resolution (a-i).}
\label{Fig:4}
\end{figure}

The transmission coefficients calculated as a function of $k_z$ (transverse wave vector along $z$-axis) for various radii of the nanorod are plotted in Fig. \ref{Fig:4}. From our previous studies \cite{Ref11,Ref12} we are able to establish that the shape of the transmission coefficient curves plotted in Fig. \ref{Fig:4}  is reasonable enough to produce satisfactory images. However, images have been observed only at the selected range of radii (filling ratios) as shown in Fig. \ref{Fig:3}. This apparent contradiction can be eliminated by studying the phase of the transmission coefficients. It was demonstrated in Ref. \cite{Ref11} that a good imaging capability requires a constant transmission phase in addition to a flat transmission coefficient magnitude over the range of transverse wave vector corresponding to the evanescent wave region.\\

\begin{figure}[htb]
\centerline{\includegraphics[width=8.70cm]{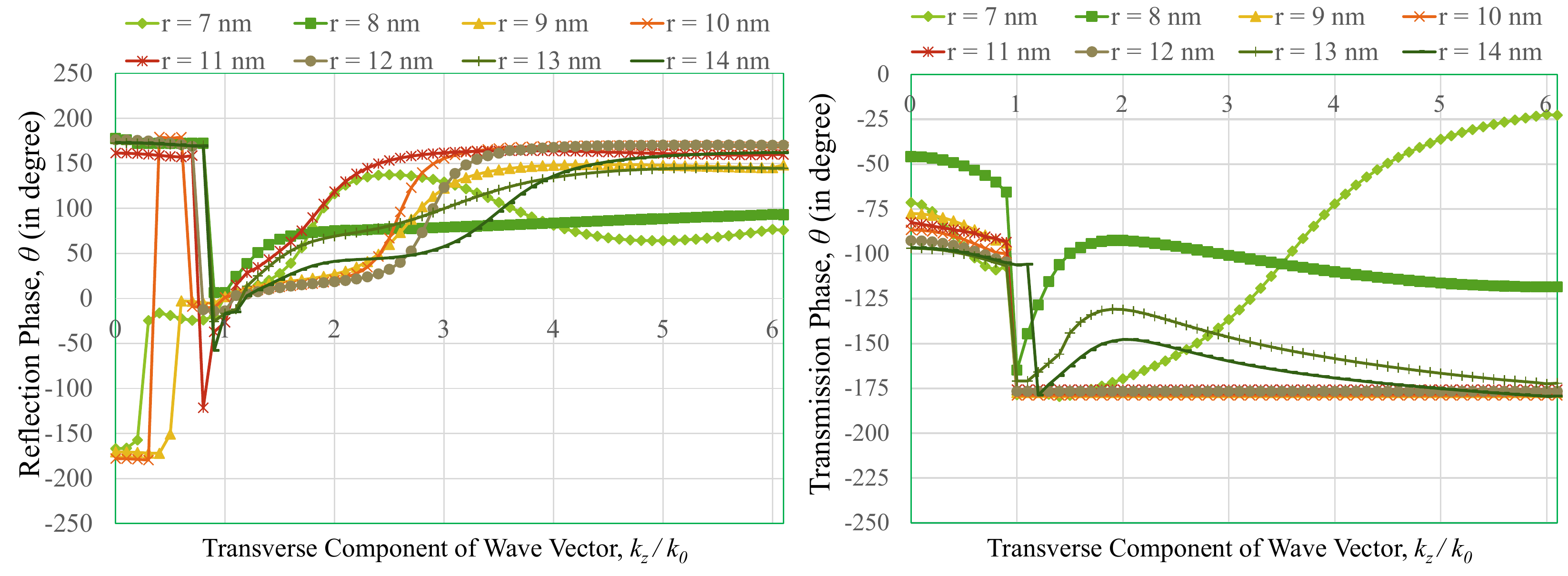}}
\caption{Phase of transmission and reflection coefficients for various radii or filling factor of the nanorod.}
\label{Fig:5}
\end{figure}

Fig. \ref{Fig:5} shows the reflection and transmission phase for different radii varying from 7 to 14nm. It is evident that rod radii from 9 to 12nm produce constant phase while the rest demonstrates significant variations over the ranges of evanescent harmonics. A constant transmission phase ensures that all evanescent harmonics propagate through the rod array and reach the image plane simultaneously. Any variation appearing in the transmission phase leads to distortion and hence a misleading image of the source.\\

From Fig. \ref{Fig:5} we see that for the rod radius of 7, 8, 13 and 14nm the phases of the transmission coefficients exhibit considerable variations and thus prevents the formation of an $E$-field distributions at the image plane which matches with that of the source plane. Note that in contrast to the phases of transmission coefficients, the reflection coefficient phases exhibits sudden jumps in the evanescent wave region. \\

\begin{figure}[htb]
\centerline{\includegraphics[width=8.70cm]{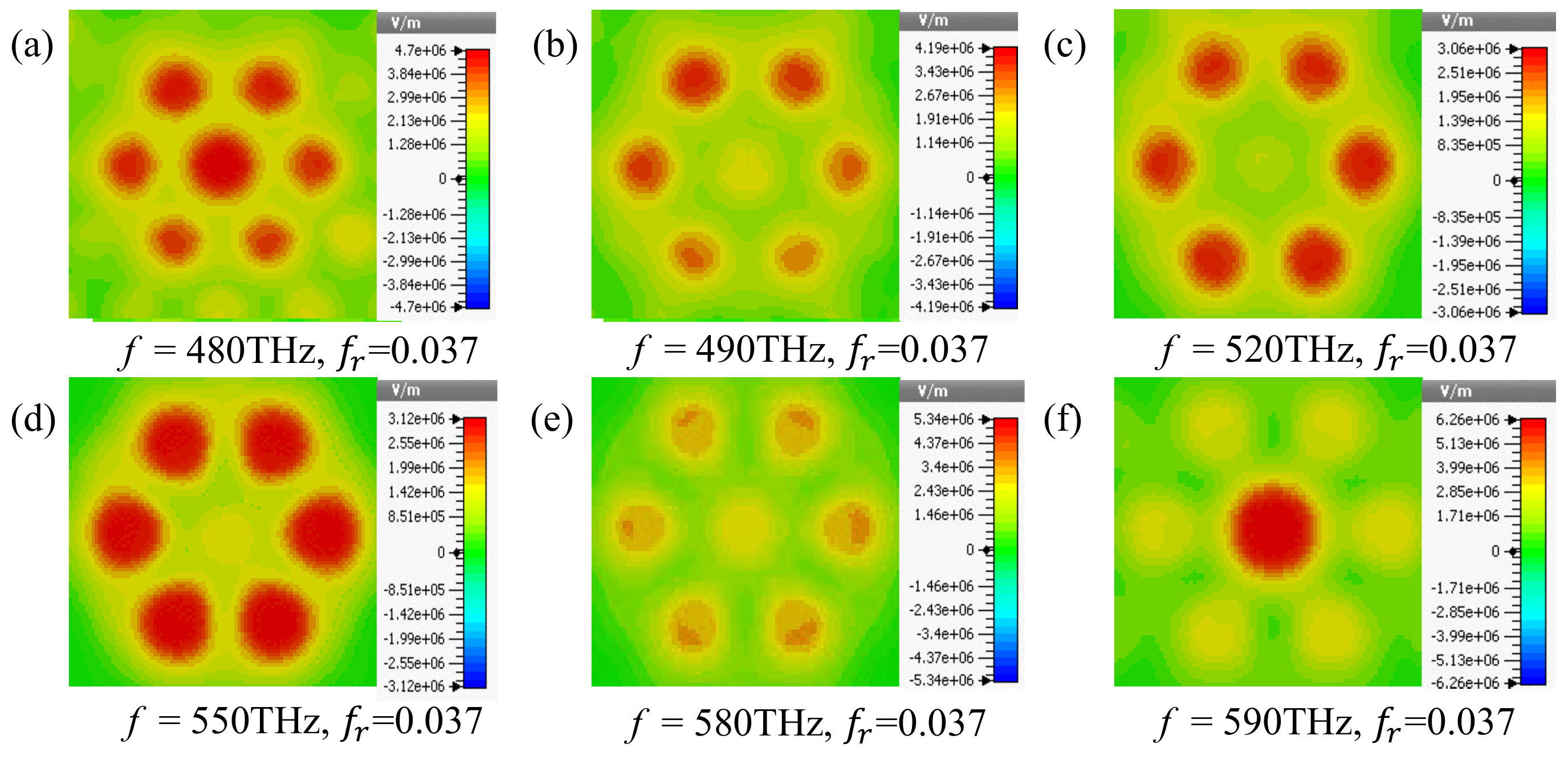}}
\caption{Image plane field distributions (a-i) for various frequencies (480THz to 590THz) for the filling ratio of 0.037 for cylindrical shaped nanorod.}
\label{Fig:6}
\end{figure}

\begin{figure}[htb]
\centerline{\includegraphics[width=8.70cm]{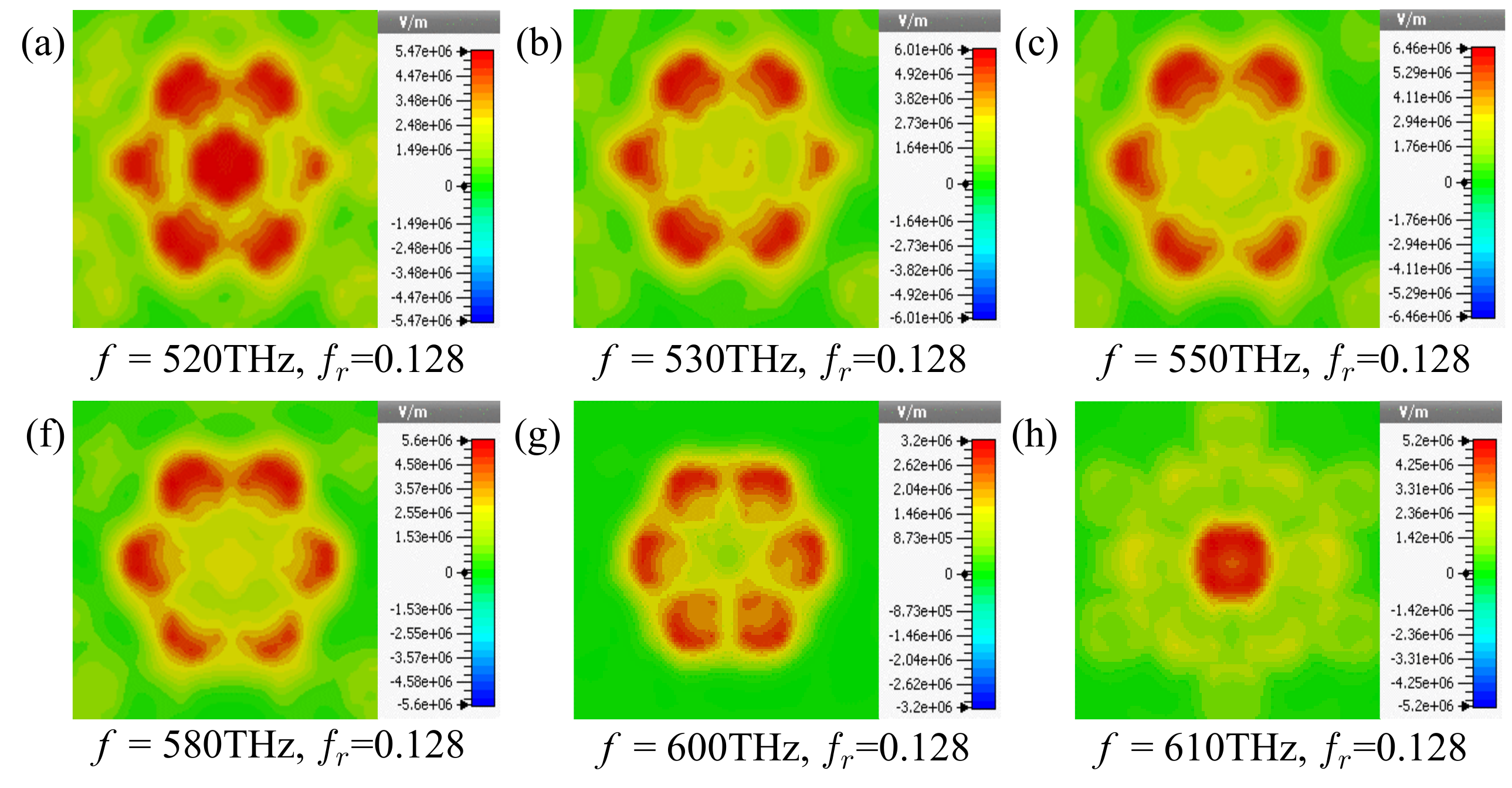}}
\caption{Image plane field distributions (a-i) for various frequencies (520THz to 610THz) for the filling ratio of 0.128 for cylindrical shaped nanorod.}
\label{Fig:7}
\end{figure}

From Fig. \ref{Fig:3}, \ref{Fig:4} and \ref{Fig:5} we can see that for 9-12nm rod radius we get identifiable images. However it is observed beyond this range  $E$-field distributions of image plane does not match with that of the source plane. The radii that failed to produce a recognizable image at 614THz have been found to work well at other frequencies as shown in Fig. \ref{Fig:6} and \ref{Fig:7}. For a rod radius of 8nm, the lens works best at 550 THz (Fig. \ref{Fig:6}), whereas, for the radius of 13nm the best image was observed at around 600THz (Fig. \ref{Fig:7}). Finally we calculated the corresponding operational bandwidth for 8nm ($f_r=0.037$) and 13nm ($f_r=0.128$) which is 16.82\% and 12.39\% respectively.

\section{Impact of filling ratios on triangular cross-sectional nanorod array}
\begin{figure}[htb]
\centerline{\includegraphics[width=8.70cm]{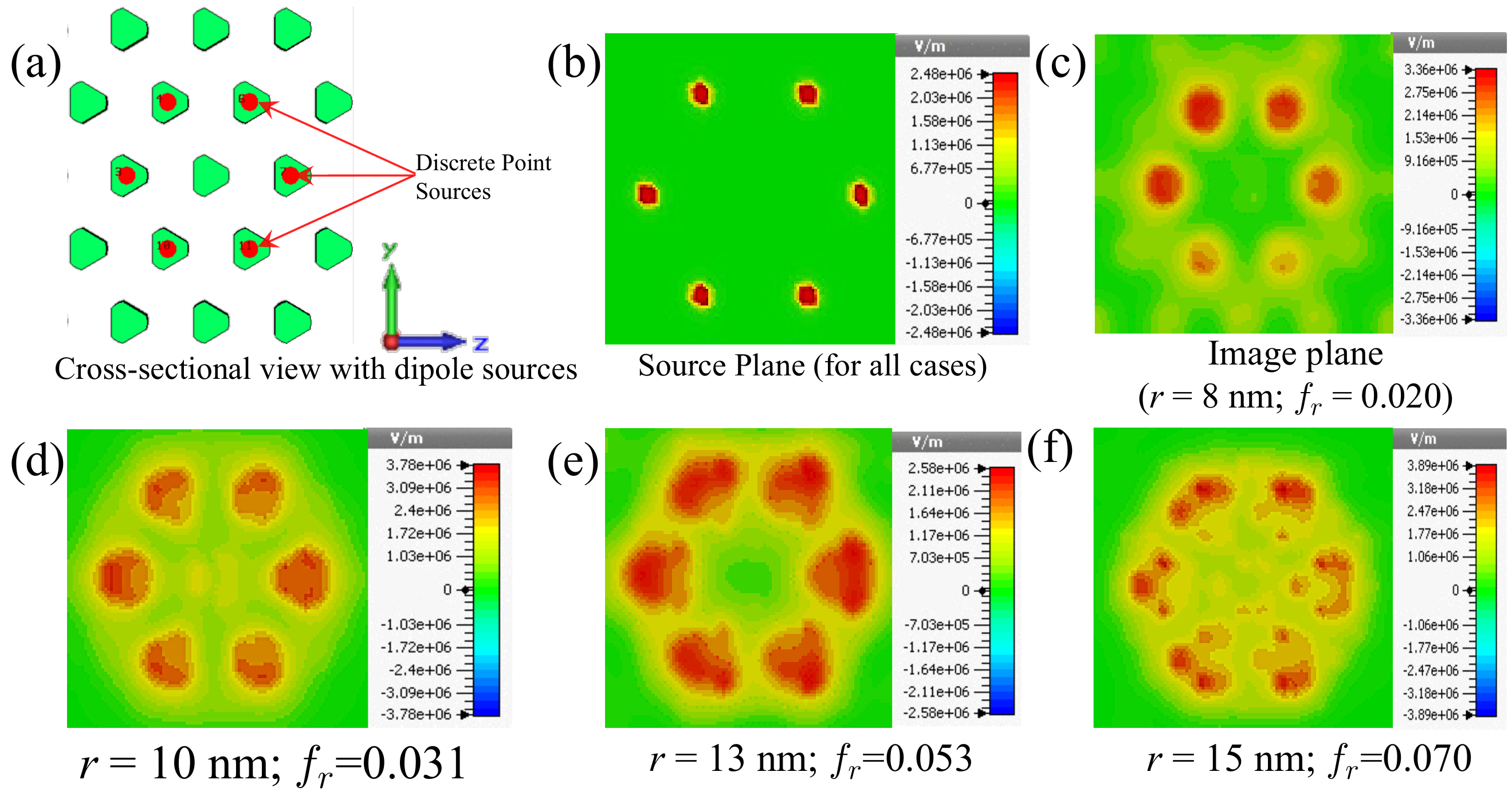}}
\caption{Source plane and image plane field distributions at 614THz for various radii of triangular shaped rod array of the rod, (a) hexagonal point source, (b) Source Plane and (c-f) image plane when $f_r$= 0.20, 0.031, 0.053 and 0.070 respectively.}
\label{Fig:8}
\end{figure}

In this section, we have made an effort to change the cross-sectional geometry of the cylindrical nanowire and observed its impact on the imaging capability of the lens. We have employed nanorod having triangular cross-section (with polished edge) and examined its performance. The $E$-field distributions, taken at the source and the image plane in 614THz demonstrate that the nano-lens offers good fidelity for the rod radii of 10nm and 13nm shown in Fig. \ref{Fig:8}. Thus it is understandable that for any radius between 10-13nm inclusive (filling ratio: $0.031 \leq f_r \leq 0.053$), the device would be operated satisfactorily. The imaging capability of the nano-lens, however, is hampered for radius below 8nm and above 15nm as the $E$-field distributions produce misleading vision of the source.\\

\begin{figure}[htb]
\centerline{\includegraphics[width=8.70cm]{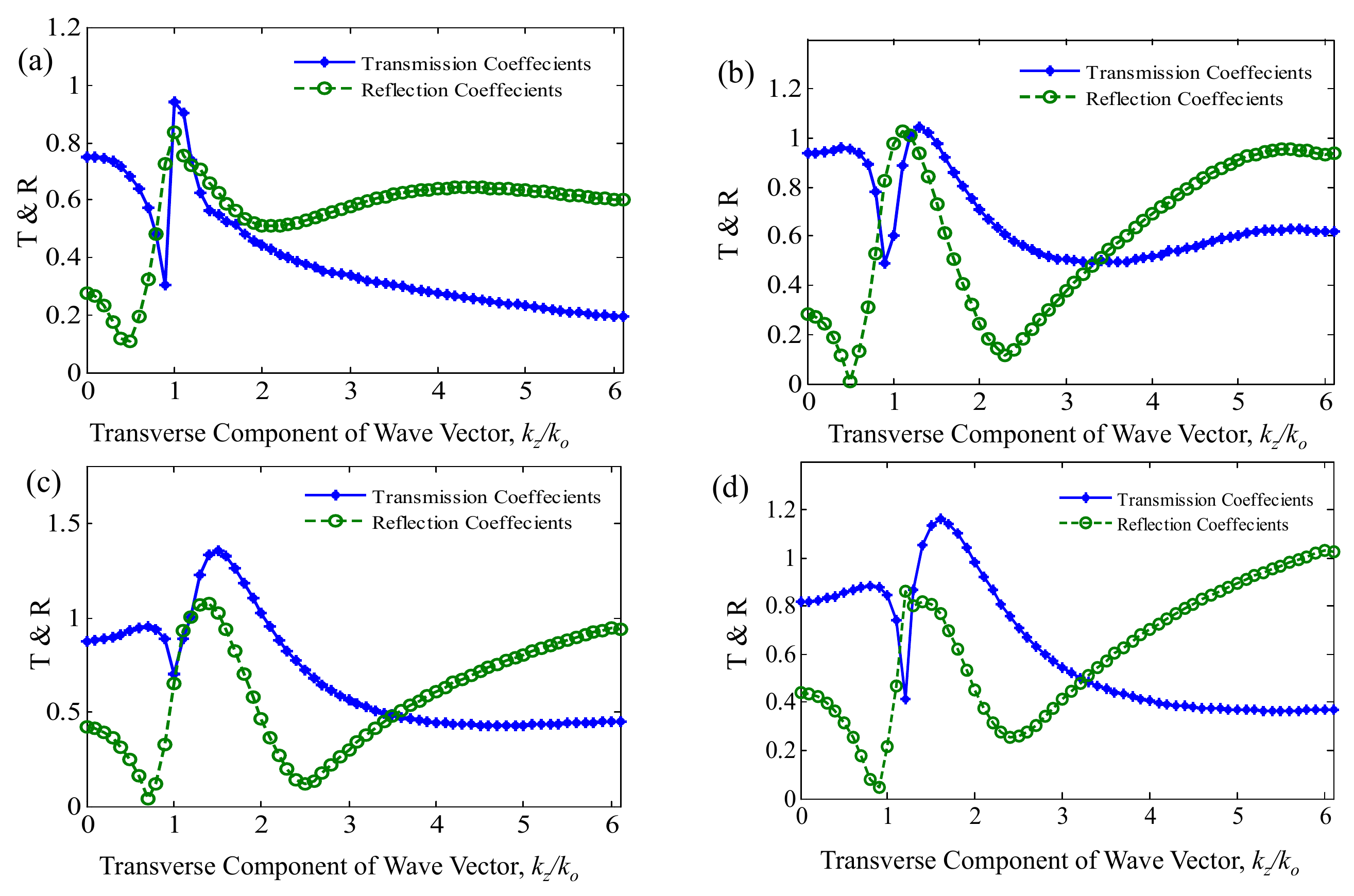}}
\caption{Transmission Coefficient (TM Modes) magnitudes at different radius of triangular shaped Ag rod at an operating wavelength of 614THz. In each case the length of each rod and the periodicity are kept same as the ref. \cite{Ref8} while the radius, $r$ =8, 10, 13 and 15nm respectively (a-d)}
\label{Fig:9}
\end{figure}

\begin{figure}[htb]
\centerline{\includegraphics[width=8.70cm]{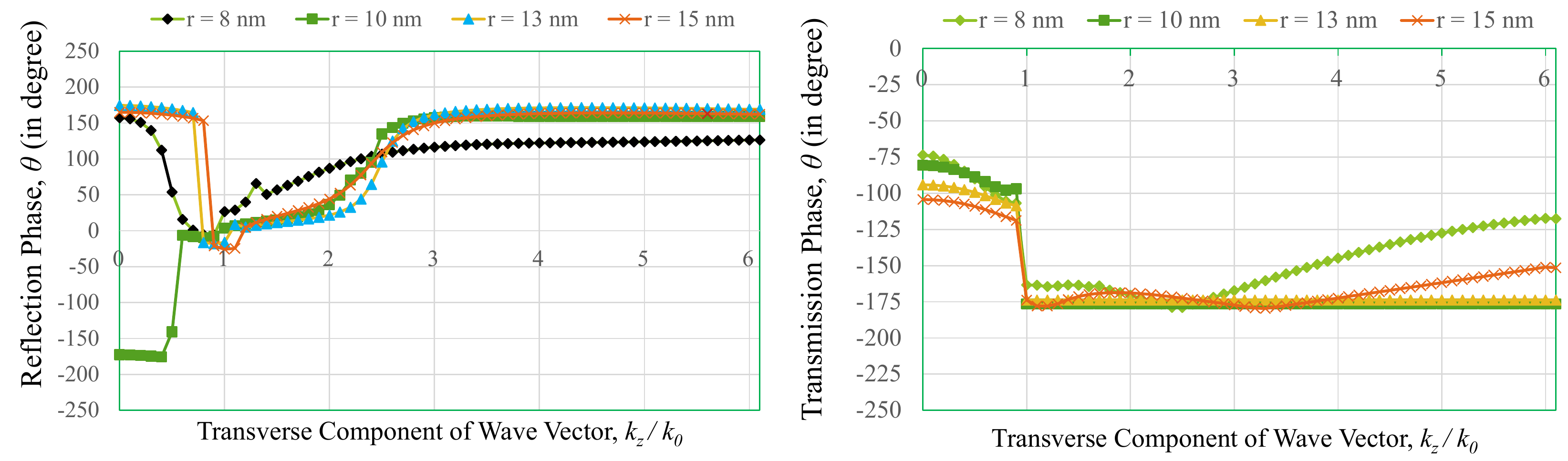}}
\caption{Phase of transmission and reflection coefficients for various radii or filling factor of the nanorod.}
\label{Fig:10}
\end{figure}

The imaging behavior of this lens can again be explained with the help of transmission characteristics. The magnitudes of transmission coefficients plotted in Fig. \ref{Fig:9} shows reasonably acceptable behavior for the radii of 8, 10, 13 and 15nm. The phases, however, exhibit superior behavior, as shown in Fig. \ref{Fig:10}, only for 10 and 13nm radii. They demonstrate non-varying characteristics that actually enable the device to function properly. The improper images of the source obtained at image planes for 8 and 15nm are attributed to the varying nature of the phases that dilute the image quality and preclude the device from functioning properly. Note that the $E$-field distribution at the image plane obtained with 15nm radius is associated with undesired fields and thus resulting in a very noisy image that is hardly recognizable, whereas, for the radius of 8nm the image is characterized by uneven field intensity for the point sources thus giving a different definition of the source than the actual one.\\

\begin{figure}[htb]
\centerline{\includegraphics[width=8.70cm]{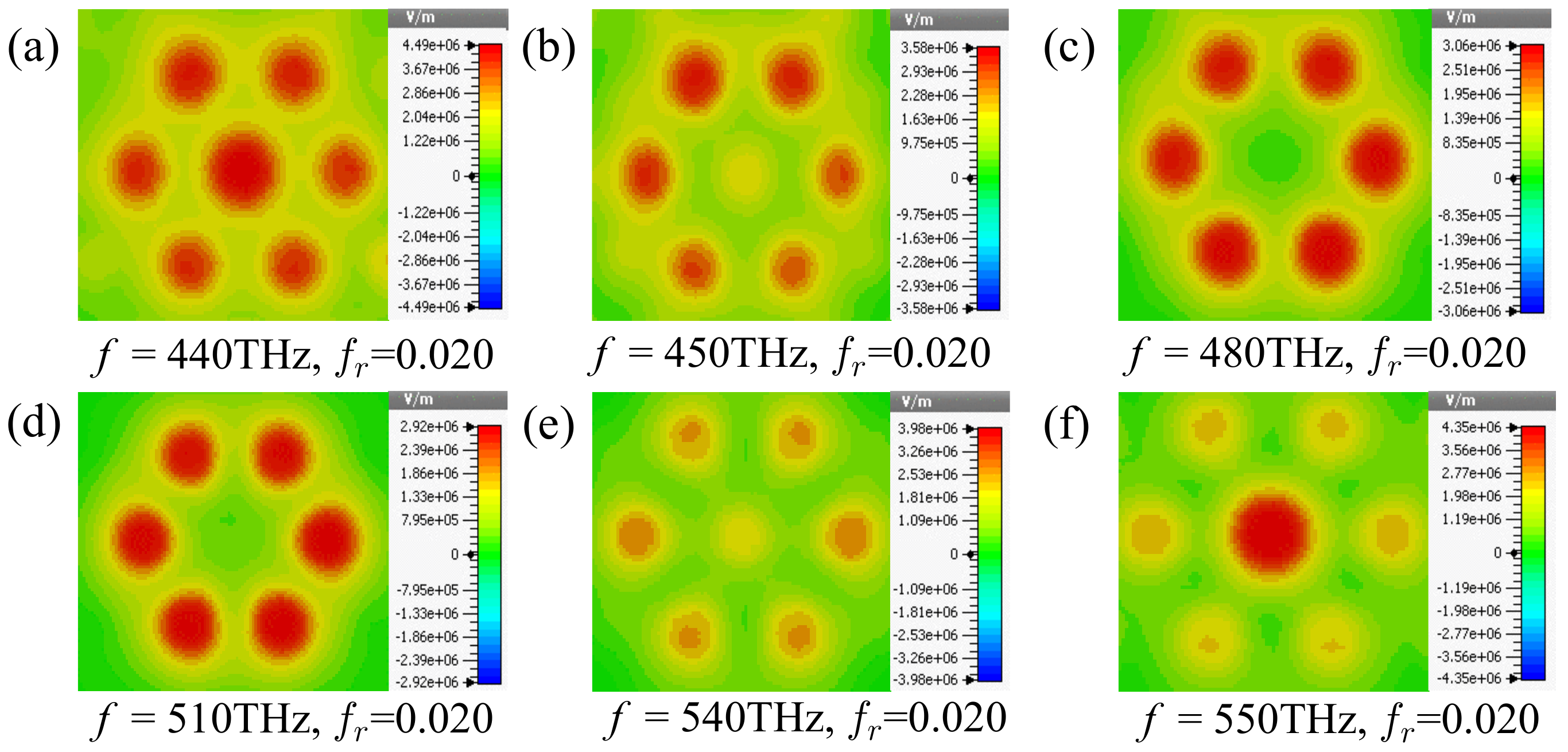}}
\caption{Image plane field distributions (a-i) for various frequencies (440THz to 550THz) for the filling ratio of 0.020 for triangular shaped nanorod.}
\label{Fig:11}
\end{figure}

\begin{figure}[htb]
\centerline{\includegraphics[width=8.70cm]{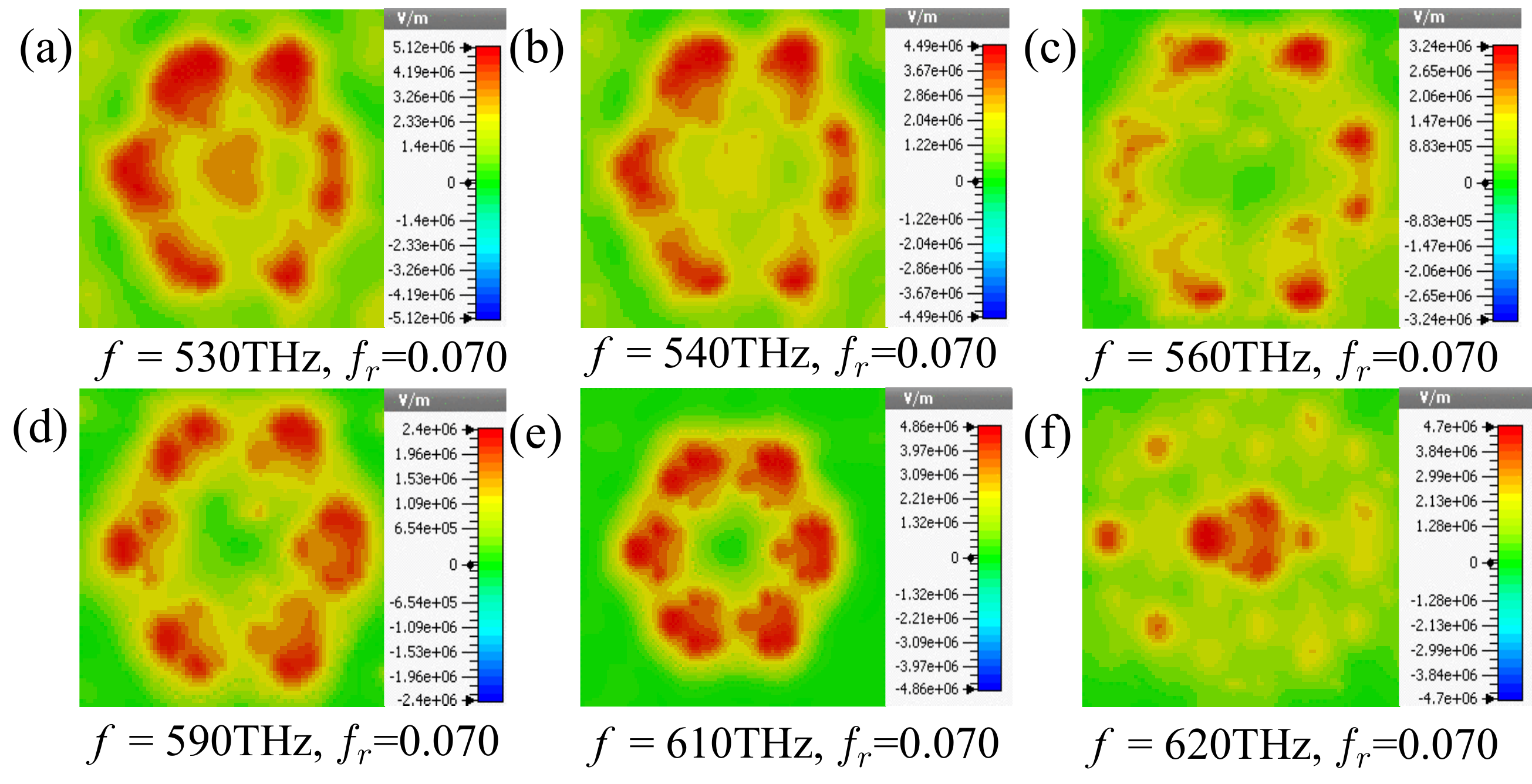}}
\caption{Image plane field distributions (a-i) for various frequencies (530THz to 620THz) for the filling ratio of 0.070 for triangular shaped nanorod.}
\label{Fig:12}
\end{figure}

Similar to the circular cross-section, we have also noticed for triangular cross-section that the filling ratios those failed to produce any image at 614THz have been found to work at other frequencies as demonstrated in Fig. \ref{Fig:11} and \ref{Fig:12}. We have summarized our studies in Table \ref{Table1} that contains information about the operating band for different radii and hence the filling ratios. Note that the quality of the image becomes poor at higher filling ratio for both the circular and triangular geometries and may be attributed to the increased coupling between surface plasmons of adjacent rods. The idiosyncratic shift of operating frequency range with the change in filling ratio can also be due to the varying degree of coupling with the changing radius. Thus the behavior of nanolens in the visible domain is much more complex than it is in the microwave regime.\\

\begin{table}
\arrayrulecolor{black!50}
\centering
\caption{Operating range and bandwidth for varying radii / filling ratios for cylindrical and triangular cross-sections.}
\label{Table1}
\begin{tabular}{ C{0.2\linewidth}  C{0.2\linewidth}  C{0.2\linewidth}  C{0.2\linewidth} }
 \toprule[1.2pt]
\multicolumn{4}{c}{\textbf{Cylindrical cross-section}}\\
 \toprule[1.2pt]
  Radius, `$r$' (\si{\nano\meter}) & Frequency range (\si{\tera\hertz}) & Center Frequency (\si{\tera\hertz}) & Bandwidth (\%) \\\toprule[1.2pt]
  8  & 490-580 & 535 & 16.82 \\ 
  9  & 590-650 & 620 & 9.68  \\ 
  10 & 550-630 & 590 & 13.56 \\ 
  11 & 590-640 & 615 & 8.13  \\ 
  12 & 560-660 & 605 & 16.53 \\ 
  13 & 530-600 & 565 & 12.39 \\ 
 \toprule[1.2pt]
 \multicolumn{4}{c}{\textbf{Triangular cross-section}}\\
 \toprule[1pt]
  8  & 450-540 & 595 & 15.13 \\ 
  10 & 530-620 & 575 & 15.65 \\ 
  13 & 540-630 & 585 & 15.38 \\ 
  15 & 540-610 & 575 & 12.17 \\ 
 \toprule[1.2pt]
\end{tabular}
\label{Table1}
\end{table}

\section{Conclusion}
We have demonstrated the impact of filling ratios on subwavelength imaging capabilities of a nanolens formed by an array of silver nanorod. We have observed that impact for two different cross-sectional geometries by analyzing image quality and calculating transmission and reflection coefficients. From our analysis we have found that having proper filling ratio is imperative to obtaining the right imaging performance of the lens. Our study in this paper confirms that the lens can be operated at different filling ratios and therefore offers a leeway from the fabrication point of view. Our study also confirms that subwavelength imaging capability of such a lens is independent of the cross-sectional geometry and thus offers another degree of freedom.

\end{document}